\documentclass[12pt,a4paper]{article}
\usepackage{latexsym}
\font\mybb=msbm10 at 12pt
\def\bb#1{\hbox{\mybb#1}}

\def\R {\bb{R}}

\begin{document}

\begin{flushright}
CERN--TH/97--93\\
{\bf hep-th/9705095}\\
May $2$nd, $1997$
\end{flushright}

\begin{center}


{\large {\bf Non-Supersymmetric (but) Extreme Black Holes, Scalar Hair and
Other Open Problems}}

\vspace{.5cm}

{\small \sl Enhanced Version of the Contribution to the Proceedings of the
  XXXIII Karpacz Winter School of Theoretical Physics {\it Duality,
    Strings and Fields}, 13-22 February, Karpacz, Poland.}

\vspace{.9cm}

{\large
{\bf Tom\'as Ort\'{\i}n}
\footnote{E-mail address: {\tt Tomas.Ortin@cern.ch}}${}^{,}$
\footnote{Address after October 1997:  
{\it IMAFF, CSIC, Calle de Serrano 
121, E-28006-Madrid, Spain}}\\
\vspace{.4cm}
{\it C.E.R.N.~Theory Division}\\
{\it CH--1211, Gen\`eve 23, Switzerland}\\
}
\vspace{.8cm}


{\bf Abstract}

\end{center}

\begin{quotation}

\small

We give a brief overview of black-hole solutions in four-dimensional
supergravity theories and their extremal and supersymmetric limits. We
also address problems like cosmic censorship and no-hair theorems in
supergravity theories. While supergravity by itself seems not to be
enough to enforce cosmic censorship and absence of primary scalar
hair, superstring theory may be.

\end{quotation}

\vspace{2cm}

\begin{flushleft}
CERN--TH/97--93\\
\end{flushleft}

\newpage

\pagestyle{plain}


\section{Unbroken Supersymmetry in Supergravity Theories}
\label{sec-unbroken}


\subsection{Bogomol'nyi Bounds and Supersymmetric configurations}

Some of the basic material in this section can also be found
Ref.~\cite{kn:GWG}.  In general, the solutions of a supergravity
(SUGRA) theory are not invariant under (local) supersymmetry (SUSY)
transformations which can be written schematically in the form

\begin{equation}
\left\{
\begin{array}{rcl}
\delta_{\epsilon} B & \sim & \epsilon F\, , \\
\delta_{\epsilon} F & \sim & \partial\epsilon + \epsilon B\, , \\
\end{array}
\right.
\end{equation}

\noindent for bosons (B) and fermions (F). Purely bosonic ($F=0$) 
configurations\footnote{These are the configurations that correspond
  to classical solutions.} which are invariant under {\it some} SUSY
transformations generated by the SUSY parameters $\epsilon_{\rm
  Killing}(x)$ are said to be supersymmetric or to have unbroken
supersymmetries. By definition $\epsilon_{\rm Killing}(x)$ satisfies
the {\it Killing spinor equation}

\begin{equation}
\delta_{\epsilon_{\rm Killing}}F \sim \partial\epsilon_{\rm Killing} 
+\epsilon_{\rm Killing} B=0  \, ,
\end{equation}

\noindent and is called {\it Killing spinor}. The Killing spinor has to be
a generator of the global SUSY algebra at infinity (i.e.~a constant
spinor in asymptotically flat spaces). A solution of the Killing
spinor equation that goes to zero at infinity simply generates SUSY
gauge transformation and is irrelevant.

Classical solutions of SUGRA theories with unbroken supersymmetries
enjoy special properties:

\begin{enumerate}
  
\item Classical supersymmetric solutions are simpler and depend on a
  smaller number of functions. Their simplicity is also due to the
  fact that:

\item In general a configuration admitting Killing spinors also admits
  Killing vectors which can be formed with the bilinear

\begin{equation}
k^{\mu}_{\rm Killing}\sim \overline{\epsilon}_{\rm Killing}
\gamma^{\mu} \epsilon_{\rm Killing}\, .
\end{equation}

\noindent $k^{\mu}_{\rm Killing}$ is timelike or null on $d=4$,
null in $N=1,d=10$ SUGRA etc.\footnote{A exception seems to be $N=1,
  d=11$ SUGRA.} The null Killing vector is associated to massless
representations of the global SUSY algebra with vanishing values of
the central charges and the associated solutions are gravitational
waves. The timelike Killing vectors are in general associated to
massive representations of the global SUSY algebra with or without
central charges and the corresponding solutions are objects like black
holes (BHs) and (M-, D-) $p$-branes.

\item Supersymmetric configurations saturate Bogomol'nyi (B) bounds
  when they are asymptotically flat and massive\footnote{Other bounds
    could be found for different asymptotics.  For instance, for
    asymptotically Taub-NUT spaces \cite{kn:KKOT}}. These bounds can
  be obtained via a Nester construction \cite{kn:WNINGH} (see also
  \cite{kn:ILPT}) or by associating the classical solution to a state
  in the quantum SUGRA theory. In the first case, certain assumptions
  concerning boundary and energy conditions have to be made. In the
  second case, one first has to make sure that such a states really
  does exist. In that case, given the action of the $N$-extended
  global SUSY algebra on the state one can derive the inequalities
  \cite{kn:WOFSZ}

\begin{equation}
M-|Z_{i}|\geq 0\, ,
\hspace{1cm}
i=1,\ldots,[N/2]\, ,    
\end{equation}

\noindent where the $Z_{i}$'s are the central charge matrix's 
complex skew eigenvalues and are complicated combinations of the
electric and magnetic charges of the vector fields in the supergravity
multiplet ({\it graviphotons)} and the constant values of the scalars
at infinity ({\it moduli}). Since a SUSY transformation on states is
of the form

\begin{equation}
\delta_{\epsilon} \sim \overline{\epsilon}\ Q\ |>\, ,  
\end{equation}

\noindent the existence of Killing spinors in the associated classical 
configuration is related to the state's annihilation by
$\,\overline{\epsilon}_{{\rm Killing}}^{\infty}\ Q\,$ which implies
that at least one of the above B bounds (say $i=1$) is saturated

\begin{equation}
M=|Z_{1}| \geq  |Z_{i}|\, ,
\hspace{1cm}
i\neq 1\, .
\end{equation}

The number of bounds saturated depends on the number of independent
Killing spinors (i.e.~the number of independent SUSY charges
$\overline{\epsilon}_{\rm Killing}^{\infty}\ Q$ that annihilate the
state. In $N=8,d=4$ SUGRA there are four $Z_{i}$'s and three
possibilities (up to permutations, which are related to duality
transformations):

\begin{equation}
\left\{
\begin{array}{rclclcl}
M & = & |Z_{1}| & \hspace{-1.5cm}\geq & |Z_{2}|,|Z_{3}|,|Z_{4}| & 
\Rightarrow & 1/8 \\
& & & & & & \\
M & = & |Z_{1}|= |Z_{2}| & \hspace{-1.5cm} \geq & |Z_{3}|,|Z_{4}| & 
\Rightarrow & 1/4 \\
& & & & & & \\
M & = & |Z_{1}|= \ldots =|Z_{4}| &  &  & \Rightarrow & 1/2  \\
\end{array}
\right.
\end{equation}

\noindent where the fraction of the total $N=8$ supersymmetries which
are unbroken is on the right. 

States that belong to the third case (which is the only in which the
saturated bound is duality-invariant and can be written as a
expression linear on the electric and magnetic charges) have maximal
unbroken SUSY and minimal mass for the given charge sector
which ensures maximal classical and quantum stability\footnote{In the
  ungauged SUGRA theories that we are considering here there are no
  particles carrying the graviphotons' electric and magnetic charges
  which are both, therefore, of topological nature.}. They are called
{\it BPS states}. {\it Vacuum states} have all supersymmetries
unbroken.

\item Supersymmetric configurations obey {\it no-force conditions} and
  multi-pole solutions describing several supersymmetric objects in
  equilibrium can be found. For instance, if we take two objects
  (which in a SUGRA theory would be extreme RN BHs) satisfying
  $M_{i}=\pm Q_{i}$ (which is the $N=2,d=4$ SUGRA B bound), wherever
  they are placed, the Newtonian and Coulombian forces cancel each
  other:

\begin{equation}
F_{ij}= -\frac{M_{i}M_{j}}{r^{2}_{ij}} + \frac{Q_{i}Q_{j}}{r^{2}_{ij}}\, ,
\end{equation}

\noindent and it is reasonable to expect that there are {\it static}
classical solutions describing them in equilibrium\footnote{If they do
  not satisfy the B bound, there are also solutions describing them in
  motion, but they are much more complicated, although of great
  interest in astrophysics.}. These solutions are the
Majumdar-Papapetrou solutions \cite{kn:MP} and describe arbitrary
number of BHs satisfying $M_{i}=\pm Q_{i}$ \cite{kn:HaHa}.

\end{enumerate}


\subsection{Supersymmetric Embeddings}

Given a solution of GR coupled to matter, we would like to known
whether it is supersymmetric or not. For that we have to be able to
identify it with a solution of a SUGRA theory, i.e.~we have to find a
solution of a SUGRA theory with the same metric\footnote{In general it
  will have different matter fields.}, or a {\it supersymmetric
  embedding} of the original solution whether or not it has unbroken
supersymmetries. If there is one, then there are many possible
embeddings for a given GR solution, due to the dualities if the SUGRA
theories. These dualities in general respect unbroken supersymmetries
and, thus, all duality-related embeddings are equally good from that
point of view. Apart from the duality-related embeddings there are
some more or less exceptional embeddings which are not duality-related
to the others and have a different number of unbroken supersymmetries.

\begin{description}

\item[Example 1] The electric RN BH solution is a solution of the Einstein
  Maxwell system

  \begin{equation}
    \label{eq:EinsteinMaxwell}
     S=\int d^{4}x\sqrt{|g|} \left[R-F^{2}\right]\, .
  \end{equation}
  
  Given that the bosonic part of the action of $N=2,d=4$ SUGRA
  coincides with the the Einstein-Maxwell action, an embedding of the
  electric RN BH solution consists in identifying the metric and the
  vector field of both theories. There is a second embedding: one can
  identify the metrics and the vector field of the Einstein-maxwell
  system with the dual vector field of the SUGRA theory. Both
  embeddings are related by electric-magnetic duality and both have
  the same number of unbroken supersymmetries ($1/2$ of the $N=2$ in
  the extreme limit.).

\item[Example 2] If we want to embed the same solution into $N=4,d=4$
  SUGRA, whose action is

  \begin{equation}
  \label{eq:N=4} 
  S=\int d^{4}x\sqrt{|g|} \left[R +2(\partial\phi)^{2} 
      +e^{4\phi}(\partial a)^{2}-e^{-2\phi}\sum_{i=1}^{6}(F^{i})^{2}
      -a\sum_{i=1}^{6}F^{i}{}^{\star}F^{i}\right]\, ,
  \end{equation}
  
\noindent we also have to satisfy the scalar equations of motion. 
In particular, the dilaton $\phi$ equation of motion

\begin{equation}
\nabla^{2}\phi \sim e^{-2\phi}\sum_{i=1}^{6}(F^{i})^{2}\, ,
\end{equation}

\noindent implies that to have no dilaton (which would change the RN metric)
we must have $\sum_{i=1}^{6}(F^{i})^{2}=0$. For this it is necessary
to have two non-trivial vector fields one with electric charge $q_{1}$ and
the other with magnetic charge $p_{3}=\pm q_{1}$. Different choices of
non-trivial vector fields are related by T~duality transformations.
Another embedding would consist of two vector fields with electric and
magnetic charge $q_{1}= \pm p_{1}\, ,\,\,q_{3}= \mp
p_{3}$\footnote{Two are necessary to satisfy the axion $a$ equation
  with vanishing axion.}.

\item[Example 3] To embed the RN BH into $N=4,d=4$ SUGRA coupled to
  six vector multiplets (the theory that one obtains when one
  compactifies $N=1,d=10$ SUGRA on $T^{6}$) we can simply embed $N=4$
  into $N=4+6V$ and so the vector field of the Einstein-Maxwell system
  is still identified with a graviphoton. There is another
  possibility: to identify this vector field with a matter vector
  field. This embedding breaks all supersymmetries and it is not
  related to the previous one by duality \cite{kn:KLOPP,kn:KhO1}.
  From the point of view of the string effective action one needs four
  vector fields with charges $|q_{1}|=|q_{2}| =|p_{3}|=|p_{4}|=|q|$.
  The two possible sets of embeddings correspond to the two possible
  relative signs $q_{1}=\pm q_{2}\, ,\,\, p_{3}=\pm p_{4}$.
  
\item[Example 4] To embed the RN BH in $N=8,d=4$ SUGRA (the theory
  that one obtains when one compactifies $N=2A, N=2B,d=10$ SUGRA on
  $T^{6}$, the effective field theories of the type II string
  theories) one can simply embed again $N=4+6V$ into $N=8$ using the
  embedding of $N=1,d=10$ into the $N=2,d=10$ theories (which consists
  in the identification of the NS-NS fields). However, now, the two
  kinds of embeddings described before turn out to be supersymmetric
  in $N=8$ (the matter vector fields of $N=4$ are graviphotons of
  $N=8$). There are now more possible embeddings too: from the string
  point of view, one needs four non-trivial vector fields to embed the
  RN solution into $N=8,d=4$ SUGRA. All that is required is that the
  four charges have the same absolute value: $|q_{1}|=|q_{2}|
  =|p_{3}|=|p_{3}|=|q|$. There are now eight different relative sign
  choices (embeddings up to T~dualities). In the extremal limit only a
  half (i.e.~four) of them are supersymmetric.
  
  The non-supersymmetric embeddings (that is, supersymmetric
  embeddings without unbroken SUSY) share many properties with the
  supersymmetric ones (after all, they have the same metric):
  Bogomol'nyi-like identities are satisfied \cite{kn:O1} and the
  solutions seem to have the same stability properties \cite{kn:D}.
  This may suggest that there is a theory with more SUSY ($N=16$?), of
  which $N=8$ is a consistent truncation, in which all of these
  embeddings are supersymmetric \cite{kn:KhO2,kn:O1}. This theory may
  have a 12-dimensional origin. Similar ideas have been suggested in
  Refs.~\cite{kn:d12}.

\end{description}

There seems to be a difference between embeddings in globally and
locally supersymmetric theories. The 't Hooft-Polyakov monopole in the
BPS limit can be embedded both in $N=2$ and $N=4$ SYM and in both
cases it has $1/2$ of the supersymmetries unbroken, i.e.~one and two
respectively. However, if we restrict ourselves to supersymmetric
embeddings with unbroken SUSY the extreme RN has $1/2$ of $N=2$ $1/4$
of $N=4$ and $1/8$ of $N=8$ unbroken SUSYs, i.e.~always only one.


\subsection{BH Thermodynamics and SUSY}


\subsubsection{$N=1,d=4$ SUGRA}

The bosonic part of this theory is GR. The B bound is equivalent to
the positivity of energy \cite{kn:WNINGH} $M\geq0$. The only static BH
solution is Schwarzschild's which depends only on $M$ and becomes
Minkowski space in the $M=0$ (extreme and supersymmetric) limit.
\footnote{Minkowski's space has all (one) supersymmetries unbroken. It
  is a vacuum of the theory. In all other cases this will also be true
  and we will not mention this fact any more.} If we associate a
quantum state to Schwarzschild's solution\footnote{It is not clear
  that this can be done.} (or assume some energy conditions) $M$ will
always be positive and the solution with $M<0$ which has naked
singularities will be excluded from the theory. SUSY seems to act here
as a cosmic censor \cite{kn:KLOPP}.

The temperature and entropy are given by

\begin{equation}
\begin{array}{rclcl}
T & = & \frac{1}{8\pi M} & 
\stackrel{\mbox{\tiny $M\rightarrow 0$}}{\mbox{\large $\longrightarrow$}} &
\infty\, , \\
& & & & \\
S & = & 4\pi M^{2} & 
\stackrel{\mbox{\tiny $M\rightarrow 0$}}{\mbox{\large $\longrightarrow$}} &
0 \, . \\
\end{array}
\end{equation}

Observe that $T$ does not go to zero in the extreme limit while it is
reasonable to say that Minkowski's temperature is zero. This is one of
many examples in which a family of metrics depending on continuous
parameters (in this case the mass $M$) has physical properties which
are not continuous functions of those parameters. The obvious reason
is that the family of metrics itself is not a continuous function (in
the space of metrics) of those parameters. Here, any metric with
$M\neq 0$, no matter how small, has a horizon and is different from
Minkowski's. The lesson to be learned in this simple example is that
the properties of Minkowski's space cannot be calculated by naively
taking the $M\rightarrow 0$ limit in Schwarzschild's.


\subsubsection{$N=2,d=4$ SUGRA}

The bosonic part of this theory is the Einstein-Maxwell system
Eq.~(\ref{eq:EinsteinMaxwell}). The B bound is $M\geq |Q+iP|=|Z|$.
The only static BH solution is RN's which depends on the ADM mass $M$,
the electric charge $Q$ and magnetic charge $P$. When the B bound is
saturated and the SUSY limit is reached the extreme limit in
which the two RN horizons coincide is also reached (extreme RN (ERN)
solution). The B bound ensures that no naked singularities will arise
from trespassing the extreme limit and, in this sense, SUSY acts again
as a cosmic censor. The ERN solution has one unbroken SUSY.

The temperature and entropy are given by

\begin{eqnarray}
T & = & \frac{1}{2\pi} \frac{\sqrt{M^{2}-|Q+iP|^{2}}}{M 
+\sqrt{M +\sqrt{M^{2}-|Q+iP|^{2}}}} 
\stackrel{\mbox{\tiny $M\rightarrow |Q+iP|$}}{\mbox{\large 
$\longrightarrow$}} 0\, , \nonumber \\
& & \\
S & = & \pi \left[ M + \sqrt{M^{2}-|Q+iP|^{2}}\right]^{2}
\stackrel{\mbox{\tiny $M\rightarrow |Q+iP|$}}{\mbox{\large 
$\longrightarrow$}} \pi M^{2} \, . \nonumber \\
\end{eqnarray}

Our experience in the $N=1$ case should prevent us from trusting the
above limits because the ERN geometry (with a single horizon) is
completely different from arbitrary close to extremality RN geometry.
IN fact, the Hawking evaporation of a non-extreme RN BH takes an
infinite time to produce an ERN BH.  It is, though, reasonable to
expect the temperature to be zero since the ERN has the lowest
possible mass in the quantum theory and should not Hawking-radiate.
With respect to the entropy, if we assume that the identification
between horizon area and entropy holds even at zero temperature, then
the above limit is also correct because the ERN BH has non-zero
horizon area. However, Euclidean semiclassical calculations gave as
result $S=0$ for the ERN BH \cite{kn:HoHaKGT}, in accordance with the
conventional content of the third law of thermodynamics \cite{kn:BCH}
and in open contrast with string state-counting results
\cite{kn:JKhM}. In Ref.~\cite{kn:Ho} and argument due to Sen is given
explaining the discrepancy and a discussion from the point of view of
the third law of thermodynamics can be found in Ref.~\cite{kn:Wa}. On
the other hand one may always take the point of view that the geometry
near the horizon is very different in the string picture
\cite{kn:BeBe} and the semiclassical approach breaks down there.  In
any case, the SUGRA result {\it is} $S=0$, the string theory
prediction is $S=\pi M^{2}$ and we should regard his difference as a
test (even if in {\it gedanken} experiments) for both theories.

What about rotating BHs? Let us go back to $N=1$. The extreme limit of
the Kerr BH is reached when $M=J$, before the supersymmetric limit
$M=0$ is reached (the angular momentum does not appear in the B bound
but it does in the extremality bound), but the supersymmetric limit
implies the extreme because $M=0$ implies $J=0$.  The extreme Kerr BH
has finite horizon area and usually this is identified with
non-vanishing entropy, although it is not clear whether from the
Euclidean semiclassical point of view this would be so. The
temperature is zero.

In $N=2$ the extreme and supersymmetric limits in presence of angular
momentum are even more different. The situation is essentially the
same but now $M\rightarrow |Q+iP|$ (we stress that $J$ does never
appear in the B bound) is always beyond the extremality limit and
produces naked singularities.

It seems that SUSY only acts as a cosmic censor in absence of angular
momentum.

There are more supersymmetric solutions in pure $N=2,d=4$ SUGRA (apart
from $pp$-waves) \cite{kn:T1}. They all belong to the IWP class
\cite{kn:IWP} and include objects with angular momentum and Taub-NUT
charge. The only ones with regular horizons (i.e.~BHs) are the ERN
(Majumdar-Papapetrou) ones \cite{kn:HaHa}.


\subsubsection{$N=4,d=4$ SUGRA}

The bosonic part of this theory is described by the action
(\ref{eq:N=4}). There are two B bounds: $M^{2}-|Z_{1,2}|^{2}\geq 0$.
Neither of them is in general duality-invariant separately and so we
take their product and divide by $M^{2}$ to get the duality-invariant
{\it generalized B bound}

\begin{equation}
\label{eq:genBbound}
M^{2} +\frac{|Z_{1}Z_{2}|^{2}}{M^{2}} -|Z_{1}|^{2} -|Z_{2}|^{2}\geq 0\, ,
\end{equation}

\noindent which is the one that should appear in the metric because the
metric is duality-invariant. The term $|Z_{1}Z_{2}|^{2}M^{-2}$ can be
identified with the charges of the scalars in regular BH solutions.

There are now two ways approaching the supersymmetric limit:
$M\rightarrow |Z_{1,2}|\neq |Z_{2,1}|$ ($1/4$ of the supersymmetries
unbroken) and $M\rightarrow |Z_{1}|= |Z_{2}|$ ($1/2$ of the
supersymmetries unbroken). To study them, it is useful to have handy
the most general supersymmetric solutions of this theory, the {\it
  SWIP} solutions \cite{kn:KO,kn:T2,kn:BKO}. They can be built by
following the recipe

\begin{enumerate}

\item Choose any two complex harmonic functions ${\cal H}_{1},{\cal
H}_{2}$

\begin{equation}
\partial_{\underline{i}}  \partial_{\underline{i}} 
\left(
\begin{array}{c}
{\cal H}_{1} \\
{\cal H}_{2} \\
\end{array}
\right) (\vec{x})=0\, ,
\end{equation}

\noindent which constitute an $SL(2,\R)$ doublet associated to S~duality.

\item Choose a set of complex constants $k^{(i)}$ satisfying

\begin{equation}
\sum_{i=1}^{6}(k^{(i)})^{2}=0\, ,
\hspace{1cm}
\sum_{i=1}^{6}|k^{(i)}|^{2}=1/2\, .
\end{equation}

These constants constitute an $SO(6)$ vector associated to T~duality.

\item Define the functions $U$ and $\omega_{\underline{i}}$ 

\begin{equation}
\left\{
\begin{array}{rcl}
e^{-2U} & = & 2\Im {\rm m} ({\cal H}_{1}\overline{H}_{2})\, ,\\
& & \\
\partial_{[\underline{i}} \omega_{\underline{j}]} & = &
\epsilon_{ijk} \Re {\rm e} 
\left({\cal H}_{1}\partial_{\underline{k}}\overline{\cal H}_{2}
-\overline{\cal H}_{2}\partial_{\underline{k}}
\overline{\cal H}_{1}\right)\, .\\
\end{array}
\right.
\end{equation}

\item In terms of these objects they can be constructed as follows:

\begin{equation}
\left\{
\begin{array}{rcl}
ds^{2} & = & e^{2U} \left(dt 
+\omega_{\underline{i}}dx^{\underline{i}} \right)^{2} 
-e^{-2U}d\vec{x}^{\ 2}\, , \\
& & \\
a + i e^{-2\phi} & = & {\cal H}_{1}/{\cal H}_{2}\, ,\\
& & \\
A^{i}{}_{t} & = & 2 e^{2U} \Re {\rm e} 
\left(k^{(i)} {\cal H}_{2} \right)\, ,\\
& & \\
\tilde{A}^{i}{}_{t} & = & -2 e^{2U} \Re {\rm e} 
\left(k^{(i)} {\cal H}_{1} \right)\, ,\\
\end{array}
\right.
\end{equation}

\noindent where $\tilde{A}^{(i)}_{\mu}$ is the dual vector potential.

\end{enumerate}

When the harmonic functions are chosen properly the SWIP solutions
describe isolated (i.e.~asymptotically flat) charged, point-like
objects\footnote{Angular momentum and NUT charge can also be included.}
that always satisfy the identity

\begin{equation}
\label{eq:genBbound2}
M^{2} +|\Upsilon|^{2} -4\sum_{i=1}^{6}|Q^{i}+iP^{i}|^{2}=0\, ,  
\end{equation}

\noindent where $\Upsilon$ is a complex combination of the scalar
charges.  This identity is the explicit form of the generalized B
bound (\ref{eq:genBbound}). These solutions represent extreme BHs and
so the supersymmetric limit is also the extremal limit.

The temperature is always zero. The area is zero when there are $1/2$
of the supersymmetries unbroken and finite when there are only $1/4$.
From the semiclassical Euclidean SUGRA point of view the entropy is
zero in the $1/4$ case (for the same reasons as in $N=2$ and in
contradiction with the stringy calculation). In the $1/2$ case the
entropy is zero in both schemes.

When matter is added to $N=4$ the rule that $1/4$ of the
supersymmetries unbroken implies regular horizon of finite area breaks
down (for instance, in the $a=1/\sqrt{3}$ dilaton BH) although from
the string theory point of view the entropy should remain finite. We
have also discussed that the extreme limit does no longer imply the
supersymmetric and the extremality limit in general.

In $N=4$ SUGRA SUSY acts as a cosmic censor only in absence of angular
momentum \cite{kn:CY}.


\subsubsection{$N=8,d=4$ SUGRA}

The most general solution is unknown. We have already discussed that
there are extremal but non-supersymmetric BH solutions in this theory.
In any case we expect that all $N=8$ SUGRA extreme BH solutions
satisfy a generalized duality-invariant B identity of the form

\begin{equation}
M^{-6}\prod_{i=1}^{4}(M^{2} -|Z_{i}|^{2})=0\, .  
\end{equation}

Many kinds of scalar charges do appear in this identity, and, as in
the $N=4$ case they are all of secondary type, i.e.~they are
completely determined by the graviphotons' electric and magnetic
charges.  As we are going to argue now in the next section, primary
charges should in some cases, and under certain assumptions, be
included in B bounds \cite{kn:AMO}.


\section{Scalar Charges versus SUSY and Duality} 
\label{sec-scalar}

Scalar charges, not being protected by a gauge symmetry, are not
conserved charges.  For minimally-coupled scalars the standard no-hair
theorems apply and any non-vanishing value implies the presence of
naked singularities. The prototype of this kind of singular solution
with non-trivial scalar hair (called {\it primary hair}) is the one
given in Refs.~\cite{kn:JNWALC} for the theory with a massless scalar
$\varphi$ and action

\begin{equation}
S= \int d^{4}x \sqrt{|g|}\ \left[R+{\textstyle\frac{1}{2}}
(\partial\varphi)^{2}\right]\, .
\end{equation}

The solutions take the form

\begin{equation}
\left\{
\begin{array}{rcl}
ds^{2} & = & W^{\frac{M}{r_{0}}-1}Wdt^{2} 
-W^{1-\frac{M}{r_{0}}}\left[ W^{-1}dr^{2} +r^{2}d\Omega^{2}\right]\, ,  \\
& & \\
\varphi & = & \varphi_{0} -\frac{{\cal Q}_{d}}{r_{0}}\ln W\, ,\\
\end{array}
\right.
\end{equation}

\noindent where

\begin{equation}
\left\{
\begin{array}{rcl}
W & = & 1-2r_{0}/r\, ,\\
& & \\
r_{0}^{2} & = & M^{2} + {\cal Q}_{d}^{2}\, .\\
\end{array}
\right.
\end{equation}

The solution is determined by three independent parameters: the mass
$M$, the scalar charge ${\cal Q}_{d}$ and the value of the scalar at
infinity $\varphi_{0}$. Only when ${\cal Q}_{d}=0$ one has a regular
solution (Schwarzschild). In all other cases there is a singularity at
$r=r_{0}$. It can be embedded in $N=4$ SUGRA identifying
$\varphi=2\phi$ in Eq.~(\ref{eq:N=4}). Observe that the above family
of solutions includes a non-trivial {\it massless} solution. Setting
$M=0$ above we find

\begin{equation}
\left\{
\begin{array}{rcl}
ds^{2} & = & dt^{2} -dr^{2} -Wr^{2}d\Omega^{2}\, ,  \\
& & \\
\varphi & = & \varphi_{0} -\ln W\, ,\\
\end{array}
\right.
\label{eq:massless1}
\end{equation}

\noindent with

\begin{equation}
\label{eq:massless2}
W= 1-\frac{2{\cal Q}_{d}}{r}\, .
\end{equation}

For non-minimally coupled scalars regular BH solutions with {\it
  secondary scalar hair} do exist (just see above). In those
solutions, the scalar (dilaton) charge is identical to a certain fixed
combinations of the other, conserved, charges (for simplicity we only
consider one $U(1)$ field):

\begin{equation}
\label{eq:dilatonhair}
{\cal Q}_{d}\sim \frac{P^{2}-Q^{2}}{2M}\, .
\end{equation}

In $N=4$ we found that the axidilaton charge in regular BH solutions
was equal to

\begin{equation}
\label{eq:upsilonhair}
|\Upsilon| = M^{-1}|Z_{1}Z_{2}|\, .  
\end{equation}

The existence of secondary hair does not preclude the existence of
primary hair. In fact, the solutions above can be interpreted in the
framework of string theory with primary but no secondary scalar hair
and there are solutions which have both kinds of hair at the same time
\cite{kn:ALC2}.

Primary scalar hair always seems to imply the presence of naked
singularities, and the no-hair theorem should maybe be called {\it
  no-primary hair theorem}.

In the standard derivations of the different B bound formulae only
conserved electric and magnetic charges appear and only when all the
scalar hair is secondary and given by
Eq.~(\ref{eq:dilatonhair},\ref{eq:upsilonhair}) one can derive the
generalized B bounds of the previous section in which the scalar
charges appear.

We are going to argue, however, that {\it primary scalar hair} should
be incorporated into the generalized B bounds.

Let us consider a simple example: Schwarzschild's solution (given
above just by setting ${\cal Q}_{d}=0$). This solution has no unbroken
supersymmetries, which can be understood in terms of non-saturation of
the B bound ($M \geq 0$). A Buscher T~duality transformation in the
time direction preserves the SUSY properties and asymptotic behavior
of the solution giving new asymptotically flat solution with no
unbroken supersymmetries: precisely the massless solution with only
primary scalar hair written above in
Eqs.~(\ref{eq:massless1},\ref{eq:massless2}).  It is easy to check
that this solution admits no $N=4$ Killing spinors and so it has no
unbroken supersymmetries\footnote{The dilatino SUSY transformation
  rule would be equal to $\delta_{\epsilon}\lambda^{I}\sim
  \not\!\partial\hat{\phi}\epsilon^{I}$ which only vanishes for
  $\epsilon^{I}=0$. ($I$ is an $SU(4)$ index here).}. However, the
fact that this solution has no unbroken supersymmetries would not have
been clear from the B bound point of view , had we used the
once-standard form in which primary hair should not added to it, since
its mass and all the other conserved charges are zero, meaning that
the bound would be trivially saturated.

All that happened in this transformation is that the mass $M$, which
does appear in the B bound has completely transformed in primary
dilaton charge ${\cal Q}_{d}$ which in principle does not.

It is clear that to reconcile these two results one has to admit that
the generalized B bound formula Eq.~(\ref{eq:genBbound2}) does apply
to all kinds of scalar charge and not only to the secondary-type one.
Only in this way becomes consistent the invariance of the B bound with
the covariance of the Killing spinor equations under T~duality.

Although our reasoning is completely clear when we look on specific
solutions one should be able to derive B bounds including primary
scalar charges using a Nester construction based on the SUSY
transformation laws of the fermions of the supergravity theory under
consideration. To be able to do this one has to be able to manage more
general boundary conditions including the seemingly unavoidable naked
singularities that primary hair implies.

Although we have kept this discussion strictly four-dimensional it is
easy to generalize these arguments to higher dimensions. In fact,
solutions generalizing the one above to higher ($d$) dimensions can be
straightforwardly found
 
\begin{equation}
\label{eq:higherALC}
\left\{
\begin{array}{rcl}
ds^{2} & = & W^{\frac{M}{r_{0}}-1}W dt^{2} 
-W^{\frac{1}{d-3}\left(1-\frac{M}{r_{0}}\right)} 
\left[ W^{-1} d\rho^{2} +\rho^{2}d\Omega^{2}_{(d-2)}\right]\, , \\
& & \\
\phi & = & \phi_{0} 
+\frac{{\cal Q}_{d}}{r_{0}}\ln W\, .
\end{array}
\right.
\end{equation}

\noindent where

\begin{equation}
W = 1 - \frac{2r_{0}}{\rho^{d-3}}\, ,
\end{equation}

\noindent and now

\begin{equation}
r_{0}^{2} = M^{2} + 2\left(\frac{d-3}{d-2}\right){\cal Q}_{d}  \, .
\end{equation}

For ${\cal Q}_{d}=0$ we recover the $d$-dimensional Schwarzschild
solution. In all other cases we have metrics with naked singularities
either at $\rho=0$ or $\rho^{d-3}=2r_{0}$.

Further examples can be found in Ref.~\cite{kn:AMO}. It seems that
most massless BHs found in string theory are T~dual to extremal BHs
with vanishing dilaton.

It is not clear whether it should be possible to find supersymmetric
solutions which saturate the generalized B bound with primary scalar
charges. It seems that, although the B {\it bound} should include the
primary scalar hair, the saturation of the bound is always reached
with secondary scalar hair only, but there is not proof of this fact.


\section{Cosmic Censorship, No-Hair Theorems and String Theory} 
\label{sec-stringtheory}

We have seen that (classical) SUSY seems to act as a cosmic censor in
general only in static situations. Solutions that saturate the B
bound and have angular momentum have naked singularities but are allowed
by SUSY.  In Ref.~\cite{kn:O2} it was observed that SUSY allows for
massless solutions that can be interpreted as being made of
constituents with positive and negative mass. SUSY does not forbid the
presence of negative mass as long as the total mass is not negative.

Something similar happens with primary scalar hair: SUSY does not
constrain its existence.

It is precisely here where a good quantum gravity theory should make
an improvement. In particular, once identified the elementary degrees
of freedom and rules under which they can be combined, such a theory
should predict that objects with naked singularities cannot be built
within its framework. With respect to the angular momentum problem, at
least, this seems to be a success of string theory. In SUGRA, if we
start with an ERN BH, the rules seem to allow us to add angular
momentum while keeping the B bound saturated (by diminishing the
mass). This is not possible in string theory \cite{kn:M} where it
seems that we can only add angular momentum by adding fermions
increasing at the same time the mass so the B bound is no longer
saturated but the extremality bound is exactly saturated and never
trespassed.

With primary scalar hair the situation is not so clear. In fact, one
could argue that there are no sources within string theory for
primary scalar hair and, as such, it should be impossible to build
string theory solutions with primary scalar hair. However, it also
looks difficult in string theory to identify the source of the mass
when the B bound is not saturated and the mass is not entirely
determined by the electric and magnetic charges (just like the
secondary hair).  It can be argued that the duality transformation
that takes us from Schwarzschild's BH (for which a string theory model
is still lacking) to the massless purely scalar solution, and which we
have use heuristically, is not a good string symmetry, but that would
not change the fact that the massless purely scalar solution is a
solution of the low-energy effective action at least as good as
Schwarzschild's. Clearly more work on this area is needed to clarify
these crucial issues.


\section*{Acknowledgments}

The author would like to thank the organizers of this very interesting
Karpacz school for their invitation to participate and
M.M.~Fern\'andez for her support.


\end{document}